\documentclass[amsmath,twocolumn,superscriptaddress,prl,aps]{revtex4-1} 
\usepackage{amsthm,amsfonts,graphicx,verbatim, color}
\usepackage{bm}
\usepackage[utf8]{inputenc}
\let\oldmarginpar\marginpar
\renewcommand\marginpar[1]{\-\oldmarginpar[\raggedleft\footnotesize #1]%
{\raggedright\footnotesize #1}}

\newcommand{\be}{\begin{equation}}
\newcommand{\ee}{\end{equation}}
\newcommand{\bea}{\begin{eqnarray}}
\newcommand{\eea}{\end{eqnarray}}

\renewcommand{\epsilon}{\varepsilon}
\renewcommand{\vec}[1]{{\bf #1}}

\def\beq{\begin{equation}}
\def\eeq{\end{equation}}
\def\bea{\begin{eqnarray}}
\def\eea{\end{eqnarray}}

\begin{document}

\title{Weyl and Dirac Loop Superconductors}
\author{Rahul Nandkishore}
\affiliation{Department of Physics, University of Colorado-Boulder, Boulder, CO 80302}
\affiliation{Center for Theoretical Quantum Matter, University of Colorado-Boulder, Boulder, CO 80302}
\begin{abstract}
We study three dimensional systems where the parent metallic state contains a loop of Weyl points. We introduce the minimal $\vec{k} \cdot \vec{p}$ Hamiltonian , and discuss its symmetries. Guided by this symmetry analysis, we classify the superconducting instabilities that may arise. For a doped Weyl loop material, we argue that - independent of microscopic details - the leading superconducting instability should be to a fully gapped chiral superconductor in three dimensions- an unusual state made possible only by the non-trivial topology of the Fermi surface. This state - which we dub the `meron superconductor' -  is neither fully topological nor fully trivial. Meanwhile, at perfect compensation additional states are possible (including some that are fully topological), but the leading instability depends on microscopic details. We discuss the influence of disorder on pairing. In the presence of a spin degeneracy (`Dirac loops') still more complex superconducting states can arise, including a `skyrmion' superconductor with topological properties similar to superfluid He III-B, which additionally breaks lattice rotation symmetry and exhibits nematic order.
\end{abstract}
\maketitle

The discovery of Dirac semimetals such as graphene, topological insulator surface states, and three dimensional Dirac and Weyl semimetals has provided an exciting new platform for solid state physics \cite{CastroNeto, HasanKane, QiZhangRMP, Hosur, Bernevig, TaAs}. These materials, which have properties intermediate between metals and insulators, exhibit qualitatively new phenomena, such as Klein tunneling and a half integer quantum Hall effect. Recently a new chapter has been opened with the theoretical discovery of `nodal loop semimetals' - three dimensional systems with a Fermi surface (FS) that at perfect compensation contains a loop of Dirac nodes, and away from perfect compensation is a torus \cite{Carter, Chen, Schaffer, Kim, Mullen, Zeng, Weng, Yu, Xie, Hasan, Schnyder, FangFu}. Spin degeneracy may be either present (Dirac loop semimetal) of absent (Weyl loop semimetal). These new classes of semimetal combine a vanishing low energy density of states and spinor wavefunctions with a FS of nontrivial topology, and may support qualitatively new phenomena. For example, it has been argued \cite{Mullen} that a time varying electric field can produce a three dimensional quantum Hall effect. 

Superconducting states in Dirac semimetals are of particular interest, for both fundamental reasons and technological applications. For example, superconducting states on the surface of a topological insulator have vortices that trap Majorana zero modes \cite{FuKane}, while doped graphene may support chiral superconductivity \cite{NandkishoreLevitovChubukov}. Nodal loop semimetals, with their unusual FS topology, may host particularly exotic and interesting forms of superconductivity. However, the superconducting states in these systems have neither been explored nor classified, although it has been noted that the superconducting transition at perfect compensation proceeds via an interesting critical point \cite{SenthilShankar}. 

In this Letter we explore and classify the superconducting instabilities of Weyl and Dirac loop semimetals
The focus is on superconducting states that might arise as weak coupling instabilities of the (topologically non-trivial) Fermi surface - a more restrictive condition than the usual classifications (e.g. \cite{SchnyderRyu}). We introduce a simple linearized $\vec{k} \cdot \vec{p}$ Hamiltonian to describe the system, and discuss the symmetries.  The analysis is developed both for the minimal model of a Weyl loop semimetal (with no spin degeneracy) as well as for a Dirac loop semimetal with full spin $SU(2)$ symmetry.  We classify superconducting states according to symmetry and the homotopy of the map to the gap function manifold.   Non-trivial superconducting states made possible by the non-trivial FS topology include fully gapped chiral superconducting states in three dimensions, and fully gapped nematic spin triplet states where the $\vec{d}$ vector has a skyrmion texture around the FS and spontaneously breaks rotation symmetry. The analysis opens up a new route to achieving exotic correlated states by exploiting unusual Fermi surface topologies. 

We begin by introducing the minimal low energy $\vec{k} \cdot \vec{p}$ Hamiltonian describing a non-interacting and non-dispersive Weyl loop. The minimal model has a FS that consists of a single Weyl loop at perfect compensation (for an discussion of material properties needed to stabilize this, see e.g. \cite{FangFu}). We choose our axes such that the loop is in the xy plane, and is a perfect circle. It then becomes natural to use cylindrical co-ordinates $(p_x, p_y, p_z) = (p_{\perp} \cos \theta, p_{\perp} \sin \theta, p_z)$. In cylindrical co-ordinates, the (first quantized) low energy non-interacting $\vec{k} \cdot \vec{p}$ Hamiltonian takes the form \cite{YongBaek}
\begin{figure}
\includegraphics[width = \columnwidth]{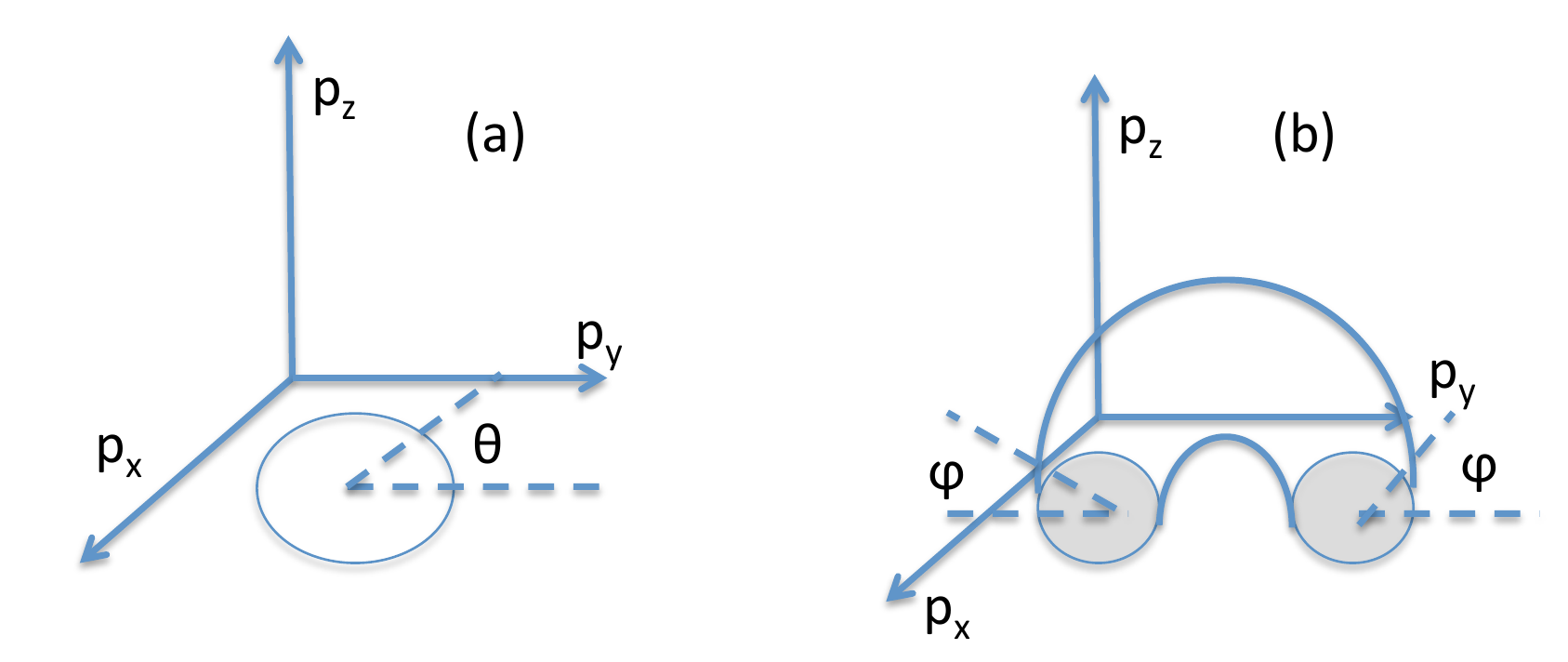}
\caption{\label{Fig} (a) The Fermi surface (FS) of a perfectly compensated Weyl loop material is a circle in the $x-y$ plane, parametrized by an angle $\theta$. (b) For a doped Weyl loop material the FS is a torus, parametrized by $\theta$ and a second angle $\varphi$, defined such that $\vec{p} \rightarrow - \vec{p}$ takes $\theta \rightarrow \theta + \pi$ and $\varphi \rightarrow - \varphi$.}
\end{figure}
\begin{equation}
\hat H_0 = \frac{p_{\perp}^2-p_F^2}{2m} \sigma_1 + v_z p_z \sigma_2 
\end{equation}
where $p_F$ is the radius of the loop, and $\sigma_{1,2}$ are the first and second Pauli matrices respectively. This is a simplified model with $U(1)$ rotation symmetry in the x-y plane - in a typical material realization this will be broken down to the lattice symmetry under rotations about the $z$ axis. We will comment on the effect of lattice symmetries in due course. For the moment, we note that we can formally rewrite (1) as $
\hat H_0 =  v(p_{\perp}) (p_{\perp} - p_F) \sigma_1 + v_z p_z \sigma_2 $
where $v(p_{\perp}) = (p_F + p_{\perp})/2m$. Linearizing $v(p_{\perp}) \approx v_{\perp} = p_F/m$, introducing rescaled co-ordinates $p_{\perp}' = v_{\perp} p_{\perp}$ and $p_z' = v_z p_z$, and straightaway dropping the primes, we rewrite the Hamiltonian as
\begin{equation}
\hat H_0 = (p_{\perp}-p_F) \sigma_{1} + p_z \sigma_2   \label{kp}
\end{equation}
The low energy dispersion takes the form
\begin{equation}
\epsilon_{\pm}(\vec{p}) = \pm \sqrt{ (p_{\perp} - p_F)^2 +  p_z^2} = \pm p.
\end{equation}
where $p_z = p \sin \varphi$ and $p_{\perp} - p_F = p \cos \varphi$ (which defines $p$ and $\varphi$). For a perfectly compensated system (chemical potential $\mu = 0$) the FS is a ring in the xy plane with radius $p_F$. For a doped system it consists of a torus of radius $\mu$ centred on this ring.  
Note that under inversion $\vec{p} \rightarrow - \vec{p}$ we have $p \rightarrow p$, $\theta \rightarrow \theta + \pi$ and $\varphi \rightarrow  - \varphi$. 

This non-interacting Hamiltonian has several symmetries of note. Two important exact symmetries are a mirror symmetry $\mathcal{M}$ implemented by taking $p_z \rightarrow - p_z$ and simultaneously conjugating with $\sigma_1$; and an `antiunitary' symmetry we denote $\mathcal{T}$, implemented by complex conjugation and $\vec{p} \rightarrow - \vec{p}$ (for the spinful Dirac loop system, $\mathcal{T}$ also involves conjugating with $i \tau_2$, where $\tau_2$ is the appropriate Pauli matrix in spin space). Note that in the spinless case we have $\mathcal{T}^2 = +1$, and in the spinful case $\mathcal{T}^2 = -1$. In addition there are two approximate symmetries: a $U(1)$ rotation symmetry with generator $\hat L_z = i \partial_{\theta}$, and a $U(1)$ rotation symmetry \cite{NMHS} with generator $\hat J_{\theta} = \frac12 \sigma_3 + \hat L_{\theta}$, where $\hat L_{\theta} = - i \partial_{\varphi} = i [p_z \partial_{p_{\perp}} - (p_{\perp}-p_F) \partial_{p_z}]$. The $U(1)$ rotation symmetry with generator $L_{\theta}$ is an approximate symmetry which is present in our model but will be broken in any real material to a discrete crystal symmetry. It should be noted that there will be at least a two-fold rotation symmetry in the $x-y$ plane (since it is a mirror plane), and thus there will also be an inversion symmetry (implemented by $\vec{p} \rightarrow -\vec{p}$ and conjugation with $\sigma_1$). However, this is not an independent symmetry operation, but simply a combination of the `fundamental' discrete symmetries $\mathcal{M}$ with crystal rotation symmetry. Meanwhile, the $U(1)$ symmetry with generator $J_{\theta}$ is not even a symmetry of (1) - it is an emergent symmetry of the linearized low energy Hamiltonian (\ref{kp}) close to perfect compensation. 
Additionally, at perfect compensation there is a particle hole symmetry, whereby conjugation with $\sigma_3$ takes $\hat H_0 \rightarrow - \hat H_0$. This symmetry will be broken by interactions, and by doping. Finally, for a Dirac loop system (with spin degeneracy) there will be an additional spin $SU(2)$ symmetry implemented by generators $\tau_1, \tau_2, \tau_3$, where the $\tau_i$ are Pauli matrices in spin space. 

Superconductivity may be introduced (at mean field level) through a matrix valued gap function $\vec{\Delta}$. The mean field Hamiltonian in second quantized notation may then be written in the form \cite{SigristUeda}
\begin{eqnarray}
\hat H &=& \sum_{\vec{p}} \vec{\psi}^{\dag}_{\vec{p}} \left[ \hat H_0 - \mu \hat N \right] \vec{\psi}_{\vec{p}} \nonumber\\ &+& \left[\vec{\psi}^{\dag}_{\vec{p}} \vec{\Delta}({\vec{p}}) (\vec{\psi}^{\dag})^T_{-\vec{p}}  + \vec{\psi}^T_{\vec{p}} \vec{\Delta}^{\dag}(-\vec{p}) \vec{\psi}_{\vec{-p}}\right] \label{meanfield}
\end{eqnarray}
where $\vec{\psi}^{\dag}_{\vec{p}} = (\psi_{1,\vec{p}}^{\dag}, \psi_{2, \vec{p}}^{\dag})$, the superscript $T$ denotes transpose, and $\psi^{\dag}_{i, \vec{p}}$ creates a fermion with pseudospin $i$ and momentum $\vec{p}$. In (\ref{meanfield}) we have omitted a term containing gap functions but no operators, which simply shifts the ground state energy. We wish to classify the possible gap functions in terms of the symmetry and topology of their momentum and pseudospin space structure. 

For a doped Weyl loop semimetal, the FS is torus shaped and non-spin degenerate. After projecting on the relevant band (valence or conduction), the Hamiltonian collapses to 
\begin{eqnarray}
\hat H &=& \sum_{\vec{p}} \vec{\chi}^{\dag}_{\vec{p}} [ p - \mu] \vec{\chi}_{\vec{p}} +\vec{\chi}^{\dag}_{\vec{p}} \vec{\chi}^{\dag}_{\vec{-p}}\Delta(\vec{p}) _{-\vec{p}}   + h.c.  \label{meanfieldsimple}
\end{eqnarray}
where the $\chi$ are creation operators in the band basis. Note there is no spinor structure left here - $\Delta(\theta, \varphi)$ is simply a function defined on the FS, which can be classified according to how it transforms under the various symmetry operations. As a periodic function of $\theta$ and $\varphi$, $\Delta$ may be expanded as a Fourier series 
\begin{equation}
\Delta(\theta, \pi) = \sum_{l_1} e^{i l_1 \theta} \left( \sum_{l_2} \Delta_{l_1 l_2}\cos (l_2 \varphi) + \sum_{l_2 \neq 0} \tilde \Delta_{l_1 l_2} \sin (l_2 \varphi)\right),
\end{equation}
where $\Delta_{l_1 l_2}$ and $\tilde \Delta_{l_1 l_2}$ are complex scalars denoting the pairing strength in the channel with $L_{z} = l_1$, $J_{\theta} = l_2$ and mirror eigenvalue $+1$ and $-1$ respectively. Recall that Fermi statistics demand $\Delta(\vec{p}) = - \Delta(-\vec{p})$ i.e. $\Delta(\theta, \varphi) = - \Delta(\theta + \pi, -\varphi)$., which implies that $\Delta_{l_1 l_2} = 0$ when $l_1$ is even, and $\tilde \Delta_{l_1 l_2} = 0$ when $l_1$ is odd. The only pairing channels that gap out the full Fermi surface have odd $l_1$ and $l_2 = 0$. Since a fully gapped state maximises the condensation energy, the leading weak coupling instability is thus likely to be in a channel with odd $l_1$ and $l_2 = 0$, independent of microscopic details. More generally, if we remember that the $J_{\theta}$ symmetry is only approximate, this channel may weakly mix with channels with non-zero $l_2$, but (because the mirror symmetry is exact), there will be no mixing with mirror-odd $\tilde \Delta$ channels. The leading superconducting instability is thus expected to be in a channel with  
%
\begin{equation}
\Delta_{\theta, \varphi} \sim \Delta_0 \exp(\pm i l \theta) f(\varphi)
\end{equation}
with $l$ odd, and $f(\varphi)$ a positive definite even function (e.g. $f = 1$). This state gaps out the full FS, is even under mirror symmetry, odd under $\mathcal{T}$, has $L_{z} = l$. Note that this corresponds to a fully gapped {\it chiral} superconducting state in three dimensions, which spontaneously breaks time reversal symmetry. This is our first major result, and suggests that doped Weyl loop materials are natural candidates for hosting three dimensional chiral superconductivity.

We now discuss the topological properties of the above-mentioned state. The topological classification depends on the homotopy of the map from the Brilliouin zone (a three dimensional torus) to the gap function space, subject to the antisymmetry constraint $\Delta(\vec{k}) = - \Delta(\vec{k})$. This homotopy classification is determined by third homotopy group $\pi_3$ of the order parameter space \cite{MooreBalents, SchnyderRyu}. If we demand that the order parameter vanishes far from the FS (as appropriate for weak coupling), then we can introduce an unconventional topological invariant, valid only for weak coupling, which is classified by the homotopy of the map from the FS to the order parameter manifold, given by the second homotopy group $\pi_2$. For the doped Weyl loop material, the order parameter space is topologically a circle ($S^1$), since the phase is the only variable. Since $\pi_3(S^1) = 0$ and $\pi_2(S^1) = 0$, there are no true topological invariants associated with the fully gapped chiral state introduced above. One may, however, wonder if there are `weak topological invariants' associated with topological order in the xy plane. 

To address the question of weak invariants, we recall that in the Bogolioubov-deGennes theory, the wavefunction is described by a normalized two component `Nambu spinor' $(u_{\vec{p}}, v_{\vec{p}})$ with arbitrary overall phase, represented on the two dimensional `Bloch' sphere. Here $(1,0)$ denotes the North Pole, $(0,1)$ the South Pole, and $\frac{1}{\sqrt{2}}(1, e^{i\theta})$ points on the equator. Chiral phases then come in two topologically distinct forms, according to whether the map from the domain to the range of the gap function does or does not wrap the Bloch sphere (the `weak pairing' and `strong pairing' phases respectively \cite{ReadGreen, Roy1}). We further recall the relations \cite{ReadGreen}
\begin{equation}
u_{\vec{p}} = \sqrt{\frac12 ( 1 + \xi_{\vec{p}}/E_{\vec{p}})} \quad v_{\vec{p}} = - u_{\vec{p}} \frac{E_{\vec{p}} - \xi_{\vec{p}}}{\Delta_{\vec{p}}^*}
\end{equation}
where $\xi_{\vec{p}} = \epsilon(\vec{p}) - \mu$ and $E_{\vec{p}} = \sqrt{\xi_{\vec{p}}^2 + |\Delta_{\vec{p}}|^2}.$ It then follows that the spinor $(u_{\vec{p}}, v_{\vec{p}})$ is at the North Pole when $|\Delta_{\vec{p}}|$ vanishes in a particle-like region ($\xi_{\vec{p}} > 0$), and at the South pole when $|\Delta_{\vec{p}}|$ vanishes in a hole-like region ($\xi_{\vec{p}} < 0$). In the present case, $|\Delta|$ is only required to vanish at the origin $\vec{p} = 0$ and at the Brillouin zone boundary, both of which are particle-like (this follows because $\Delta_{\vec{k}} = - \Delta_{-\vec{k}}$ and thus $\Delta$ must vanish whenever $\vec{k} = - \vec{k}$). Everywhere else in the zone $\Delta_{\vec{k}}$ can be non-zero, even while taking the appropriate values on the Fermi surface. As a result, the spinor need never touch the South Pole, and thus the map is not topological. However, unlike the topologically trivial `strong pairing' phase from \cite{ReadGreen}, the map is also not deformable to the trivial map, because all points on the Fermi surface have $\xi_{\vec{p}} = 0$ and thus map to the equator (at least in the weak coupling limit where the Bogolioubov theory is appropriate). This situation is analogous to the `critical point' between weak and strong pairing discussed in \cite{ReadGreen}, where the gap function is constrained to wrap at least half of the Bloch sphere, and is reminiscent of `merons' in quantum Hall systems \cite{merons}, which also sweep out half the Bloch sphere. We therefore dub this state a `meron superconductor.' A further investigation of the properties of this chiral, fully gapped `meron' superconducting phase is a worthwhile topic for future work, as is the question of topologically non-trivial states stabilized by Fermi surface constraints.

We now discuss systems at perfect compensation, where the FS is a circle with co-dimension 2 such that $\pi_1$ of the order parameter space provides a topological classification of the map from the FS to the order parameter manifold. The most general order parameter may be written
 \begin{equation}
\vec{\Delta} =(d_{0, \vec{p}} + \vec{d}_{\vec{p}} \cdot \sigma) i \sigma_2,
\end{equation}
where $d_{0, - \vec{p}} = d_{0, \vec{p}}$ and $\vec{d}_{-\vec{p}} = - \vec{d}_{\vec{p}}$. This is a convenient parametrization, although the absence of $SU(2)$ pseudospin symmetry prevents us from identifying $d_0$ and $\vec{d}$ as `singlet' and `triplet' components. Now the system can support a topologically trivial superconducting state ($\vec{d} = 0$ and a non-zero, momentum independent $d_0$) which does, however, break the mirror symmetry, as well as chiral superconducting states - either $\Delta \sim d_0 i \sigma_2  \exp(i l \theta)$ with $l$ even, or $\Delta \sim (\vec{\tilde d} \cdot \sigma ) i \sigma_2 \exp(i l \theta)$ with $l$ odd, where $\vec{\tilde d}$ is a reference vector. These states are fully gapped chiral (but non-topological) states,as discussed above. All of these states break time reversal symmetry, and some break mirror symmetry as well. However, if we demand that the order parameter is even under $\mathcal{M}$ and $\mathcal{T}$, then that enforces $d_0, d_1 = 0$ and real $d_3/d_2$. Thus, at perfect compensation we can have a fully symmetry preserving state characterized by a $\vec{d}$ vector with only two non-zero components $d_2, d_3$ with real ratio i.e. a two component vector that lives in a space that is topologically $S^1$. There is then an additional topological superconductor with $Z$ classification, wherein this allowed $\vec{d}$ vector winds $n$ times as we go around the loop (i.e. the $\vec{d}$ vector describes a vortex around the FS). An example of such a state would be $d_2 = \Delta_0(\vec{p})\cos n \theta$, $d_3 = \Delta_0(\vec{p}) \sin n \theta$, with $n$ odd to satisfy the antisymmetry constraint, where $\Delta_0(\vec{p})$ is a non-negative function that is maximal on the FS, and vanishes only at $p_x = p_y = 0$ and at the zone boundary. This is a topological state characterized by a $Z$ weak invariant associated with the winding of the pseudospin $\vec{d}$ vector in the $xy$ plane. It can be thought of as a `layered' version of a two dimensional crystalline topological superconductor \cite{AndoFu}, protected by $\mathcal{M}$ and $\mathcal{T}$ symmetries. Interestingly, there exist topologically distinct continuations of the above state off FS (e.g. a `hedgehog' configuration $\vec{d}_{\vec{p}} \propto \vec{p}$) which look identical on FS, and thus will be degenerate in the weak coupling limit. We leave discussion of these weak coupling degeneracies to future work.  

All the above superconducting states are fully gapped at perfect compensation. However, for small doping away from perfect compensation these states will generically connect to states where the gap function has inter-band matrix elements. These states will thus be suppressed both by doping away from perfect compensation, and by a random chemical potential. The exceptions are states with $(d_0, d_1, d_2, d_3) = e^{i l \theta} ((\Delta_c - \Delta_v) \sin \varphi,0, (\Delta_c + \Delta_v), (\Delta_c - \Delta_v) \cos \varphi)$, where $l$ is an odd integer and $\Delta_c$ ($\Delta_v$) are momentum independent constants denoting the intra-band pairing strengths in the conduction (valence) band respectively \cite{basis}. These states have zero inter-band matrix elements, and connect smoothly onto the fully gapped chiral superconducting state in the conduction (valence) band as $\Delta_v \rightarrow 0$ ($\Delta_c \rightarrow 0$). Note that taking $\Delta_c = \Delta_v$ gives a state that respects the particle-hole symmetry of the problem, while any other choice breaks the particle-hole symmetry. We thus expect that, as in \cite{NMHS}, when the particle hole symmetry is present only in a statistical sense, the random chemical potential disorder will tend to {\it enhance} pairing in the channel $(d_0, d_1, d_2, d_3) = e^{i l \theta} (0,0,\Delta,0)$ only.

Meanwhile, for a Dirac loop material there is an additional two component spin degeneracy. For a doped Dirac loop material with torus FS and spin $SU(2)$ symmetry, it is convenient to parametrize the gap function (projected on the conduction band) as \cite{NMHS} 
\begin{equation}
\vec{\Delta} =(\Delta_{s, \vec{p}} + \vec{d}_{\vec{p}} \cdot \tau) i \tau_2,
\end{equation}
where $\Delta_s$ is the pseudospin singlet piece, and $\vec{d}$ is a three component vector corresponding to the pseudospin triplet piece of the order parameter and the $\tau_i$ are Pauli matrices in spin space. Fermi statistics demand that $\Delta_{s, \vec{p}}$ should be even and $\vec{d}_{\vec{p}}$ should be odd under $\vec{p} \rightarrow - \vec{p}$. 

Once again we consider fully gapped superconducting states, the topological properties of which are characterized by the first and second homotopy groups of the order parameter manifold. The order parameter manifold for spin singlet pairing is characterized purely by a phase (i.e. topologically $S^1$). Spin singlet superconductors thus come both in trivial form ($\Delta_s(\theta, \varphi) = constant$) and in fully gapped chiral form $\Delta_s(\theta, \varphi) = e^{i l \theta}$, where $l$ is now an {\it even} integer. This state is analogous to the chiral superconducting states previously discussed in the Weyl case.  Similar chiral superconducting states (with fixed $\vec{d}$) also exist in the triplet sector. However, there can be additional states where the $\vec{d}$ vector undergoes a topologically non-trivial winding about the FS. Such topologically nontrivial windings arise if we restrict our attention to time reversal invariant spin singlet states, such that the components of $\vec{d}$ have real ratios (`unitary' order parameter \cite{SigristUeda}). The order parameter manifold for time reversal invariant triplet pairing is then topologically $S^2$. Since $\pi_2(S^2) = Z$ there is in this case the additional possibility of superconducting states with {\it strong} topological invariants, whereby the $d$ vector effectively describes a skyrmion texture around the torus. An explicit construction of such a texture may be found in \cite{QiZhang} e.g. 
 \begin{eqnarray}
 d_2 &=& \sin (\theta - \theta_0); \quad d_3 = - \sin (\varphi - \varphi_0); \nonumber\\ 
 d_1 &=& 1 - \cos (\theta-\theta_0) - \cos (\varphi - \varphi_0)
 \end{eqnarray}
In terms of the celebrated `periodic table' (e.g. \cite{Kitaev}) this corresponds to a topological superconductor in class DIII, with properties analogous to e.g. the B phase of superfluid He-III.  Note too, however, that with a simply connected FS (as in He-III), a fully symmetry preserving topologically non-trivial texture (e.g. a `hedgehog') is a possibility, which is {\it not} the case on the torus - the `skyrmion' texture above (more specifically the position of the skyrmion core $(\theta_0, \varphi_0)$) spontaneously breaks both the $\theta$ and $\phi$ rotation symmetries. These are not exact symmetries, so there will not be Goldstone modes (the position of the skyrmion core will be locked by crystal symmetries). In particular, mirror symmetry will be preserved IFF the skyrmion core sits on the $\varphi_0 = 0$ line.  Additionally, in a real crystal certain $\theta$ positions will be privileged - those corresponding to crystal symmetry axes in the plane of the loop. The choice of $\theta_0$ will then spontaneously break the discrete lattice rotation symmetry. The skyrmion texture is thus also a type of {\it nematic} state. The properties of domain boundaries between domains of different nematic configuration is an interesting topic for future work. Finally, for Dirac loop materials at perfect compensation, there is the possibility for simultaneous windings in pseudospin as well as spin space. We defer discussion of these to future work. 

{\it Conclusions} The non-trivial FS topology in Weyl and Dirac loop semimetals provides a novel route to exotic unconventional superconducting states. A doped Weyl semimetal is expected to realize a fully gapped chiral `meron' superconductor in three dimensions, while still more exotic possibilities can arise at perfect compensation or in the presence of a spin degeneracy. 
 
{\bf Acknowledgements} I acknowledge several illuminating conversations with Joseph Maciejko, S.L. Sondhi and Rahul Roy. I also thank Titus Neupert for feedback on the manuscript.


\begin{thebibliography}{99}
\bibitem{basis}
The basis transformation between the band and Dirac bases is as in \cite{NMHS}, but recall that we have $\varphi_{-\vec{p}} = - \varphi_{\vec{p}}$ whereas \cite{NMHS} had $\varphi_{\vec{-p}} = \varphi_{\vec{p}} + \pi$
\end{thebibliography}
\end{document}